\documentclass[journal]{IEEEtran}
\usepackage{amsmath}
\usepackage{graphicx}
\usepackage{cite}

\usepackage{multirow}

\usepackage{subfig}

\begin{document}
\title{Kinetic simulation study of electron holes dynamics \\ during collisions of ion-acoustic solitons}
\author{S.~M.~Hosseini~Jenab, F.~Spanier  \thanks{S.~M.~Hosseini~Jenab and F.~Spanier are with Centre for Space Research, North-West University, Potchefstroom Campus,
Private Bag X6001, Potchefstroom 2520, South Africa (e-mail: Mehdi.Jenab@nwu.ac.za, Felix@fspanier.de).}}


\maketitle
\begin{abstract}
Ion acoustic (IA) solitons are accompanied by vortex-shaped nonlinear structures
(e.g. hollows, plateaus or humps) in the electron distribution function, 
called electron holes, portraying trapped electrons. 
These structures appear as charged flexible clouds 
(shielded by the background plasma) in the phase space
with their own inertia, depending on the number of trapped electrons.
According to simulation studies, electron holes tend
to merge in pairs until one accumulative hole remains in the simulation box.
This tendency has been analytically and qualitatively explained in the frame of the energy conservation principle. 
However, 
electron holes accompanying IA solitons should not merge
due to stability of IA solitons against mutual collisions.
In this report based on a fully kinetic simulation approach,
detailed study of the collisions of IA solitons 
reveals the behavior of electron holes under these two conflicting predictions,
i.e. stability against mutual collisions and merging tendency. 
Four main results are reported here. 
Firstly, we find that
among the three different types of collisions possible for electron holes, 
just two of them happen for electron holes accompanying IA solitons.  
We present different collisions,
e.g. two large/small and large versus small holes,
to cover all the these three different types of collisions.
Secondly, we show that 
although electron holes merge during collisions of IA solitons,
the stability of IA solitons 
forces the merged hole to split and form new electron holes.
Thirdly, we reveal that holes share their trapped population during collisions.
Post-collision holes incorporate 
some parts of the oppositely propagating before-collision holes.
Finally, it is shown that the newly added population of trapped electrons 
goes through a spiral path inside the after-collision holes hole because of dissipative effects.
This spiral is shown to exist in the early stage of the formation of the holes in the context of the IA soliton dynamics.

\end{abstract}


\section{Introduction} \label{Introduction}
\IEEEPARstart{I}{on} acoustic (IA) solitons have been proven to exist 
in experimental\cite{ikezi1970formation,ikezi1973experiments,nakamura1996experiments,nakamura1982experiments},
theoretical\cite{Washimi1966996,kakutani1968reductive,Gardner19671095,taniuti1968reductive},
and simulation studies\cite{Zabusky1965}.
Both \textbf{reduced fluid (KdV-based \cite{schamel_4} and fluid \cite{Kakad2013,Sharma2015})}
and kinetic (PIC \cite{Kakad20145589, Qi20153815} and Vlasov \cite{jenab2017overtaking, jenab2017pre})
approaches have been utilized to 
study these nonlinear structures.
They are proposed to describe electrostatic solitary waves (ESWs) in the broadband electrostatic noise (BEN) 
observed by different satellites (e.g., Polar\cite{franz1998polar}, GEOTAIL \cite{matsumoto1994electrostatic,kojima1997geotail}, 
FAST \cite{catte1151998fast}, and Cluster \cite{pickett2003solitary,hobara2008cluster,pickett2004solitary})
in various regions of the Earth's magnetosphere.
They show (like other types of solitons) an extraordinary property: 
when they collide with each other, they come out of the collision intact. 

Washimi and Taniuti \cite{Washimi1966996} showed that IA solitons are in fact KdV solitons,
considering electrons as a Boltzmann's fluid (i.e. inertialess) and ions as a cold fluid. 
The dependencies of different aspects of KdV solitons on each other are as follows:
\begin{align*}
 M-1 = \delta n/3,  \ \
 D = \sqrt{6/\delta n}.
\end{align*}
This is known as nonlinear dispersion relation (NDR)
in which width ($D$) and velocity ($M$ Mach number) depend on the amplitude of density perturbation ($\delta n$).
There have been disparities between experimental results and the predictions by the NDR \cite{ikezi1970formation}. 
A few different theoretical refinements to the KdV equation have been proposed 
to address these disparities.
by adding three extra effects\cite{tran1979ion},
e.g.
finite ion temperature\cite{Tappert19722446,tagare1973effect,sakanaka1972formation},
trapped electrons\cite{tran1976trapped,schamel_3,schamel_4,ichikawa1977solitons},
and higher order nonlinearity\cite{watanabe1978interpretation,h1976contribution}. 

\begin{figure}[!t]
  \centering
  \subfloat{\includegraphics[width=0.5\textwidth]{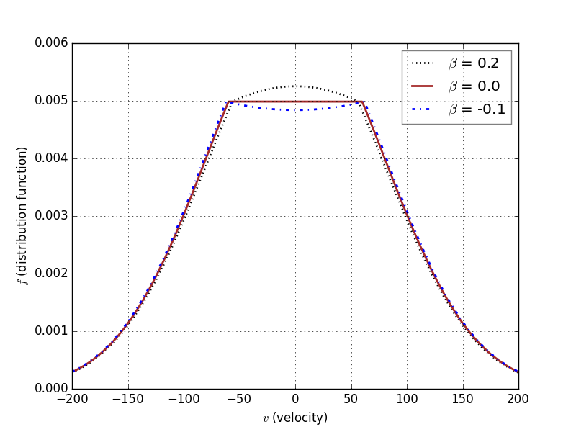}}
  \caption{Trapped electrons distribution function around $v_0=0$ 
  appears as a hollow ($\beta<0$), a plateau ($\beta=0$) and a hump($\beta>0$).}
  \label{DF_Beta}
\end{figure}

Schamel\cite{schamel_3,schamel_4} proposed a modified KdV (mKdV) equation to include the electron trapping effect into the nonlinear fluid theory approach. 
The mKdV equation possesses an extra nonlinearity caused by trapped electrons. 
This new nonlinearity is controlled by the so called \textit{trapping parameter} ($\beta$).
This parameter describes the distribution function around the soliton velocity in the phase space. 
Fig.\ref{DF_Beta} shows that the distribution function of trapped particles 
can take three different types of shapes based on the value of $\beta$, 
namely \textit{hollow} ($\beta<0$), \textit{plateau} ($\beta = 0$) and \textit{hump} ($\beta>0$).

It is shown that electron holes tend to merge in pairs in simulation studies of 
beam-plasma interactions \cite{berk1970phase,omura1996electron},
nonlinear Landau damping (Bernstein-Greene-Kruskal -BGK- modes)\cite{ghizzo1988stability}, 
and even in electron holes accompanying solitary waves in pair plasmas \cite{eliasson2005solitary}.
The merging of holes continues in pairs until the system reaches one single hole,
which has been reported to be stable in both simulation and laboratory studies \cite{berk1970phase}. 
Krasovsky \textit{et. al.} \cite{krasovsky1999interaction,krasovsky1999interaction_small,krasovsky2003electrostatic}. 
have developed a model based on the energy conservation principle to describe this phenomenon qualitatively. 
They have suggested that based on the size of electron holes, three different types of collisions might occur. 
Furthermore, they have suggested \textit{``merging condition''} which determines 
if electron holes merge during collisions.
It is also reported that electron holes, in BGK mode simulations, 
exert central force on each other even if they do not merge \cite{ghizzo1987bgk}.

The merging tendency of electron holes and the stability of IA solitons against successive mutual collisions present 
two conflicting scenarios for a collision of electron holes accompanying IA solitons. 
On one hand,
electron holes are shown to be prone to merging when they are large enough 
to have considerable overlapping in the velocity direction.
In other words, when they satisfy the \textit{``merging condition''} 
(see Sec. \ref{types_collisions} for definition of this condition). 
On the other hand,
IA solitons survive (theoretically infinite number of) mutual collisions and so do the electron holes accompanying them.
This stability is independent of their sizes, i.e. if the merging condition applies.

The goal of this study is twofold:
\begin{enumerate}
 \item Testing the three different types of collisions possible for electron holes
in the context of electron holes accompanying IA solitons. 
This is carried out by simulating collisions of IA solitons with different sizes.
 \item Examining the dynamics of electron holes accompanying IA solitons when
 they are large enough to meet the merging condition.
\end{enumerate}

The simulation approach utilizes the Vlasov equation to follow the dynamics of both plasma constituents (e.g. electrons and ions) 
coupled via Poisson's equation. 
Note that simulations presented here are in electrostatic regime. 
Collisions between IA solitons with different sizes and trapping parameters are carried out and three ones are reported here,
e.g. two small, large versus small and two large IA solitons. 
However, only head-on collisions are analyzed here and the overtaking collisions remain out of the scope of this paper. 
Furthermore, collisions are compared to the different types of collisions suggested by Krasovsky \textit{et. al.}
\cite{krasovsky1999interaction,krasovsky1999interaction_small,krasovsky2003electrostatic}.

Since the exact NDR equation
(i.e. size, velocity and shape of the IA solitons)
still stays inconclusive, IA solitons are self-consistently excited in this study. 
Any large compressional pulses in the plasma should break into a number of IA solitons,
as it has been reported in 
numerous experimental and simulation studies\cite{ikezi1970formation,ikezi1973experiments,Kakad2013}.
Mathematically, this phenomenon is explained based on N-solitons solution suggested by Hirota \cite{Hirota1971,Hirota1972}.
This phenomenon, known as chain formation, 
is employed here to produce self-consistent IA solitons.

Chain formation and collision of IA soliton have been studied through fluid simulations\cite{Kakad2013,Sharma2015};
however, they are not able to present the holes in the phase space and hence overlooked the kinetic details.
PIC\cite{Kakad20145589} and hybrid-PIC\cite{Qi20153815} simulations have ignored 
the details of trapped electrons distribution function in their reports 
mostly due to the inherent noise of PIC which smooths out holes in the phase space.
Vlasov simulation studies of the chain formation have proved the stability of self-consistent IA solitons,
produced by this method,
against successive mutual collisions\cite{jenab2017overtaking,jenab2017pre} 
and compared them to the Schamel theory\cite{jenab2016IASWs} .

\section{Basic Equations and Numerical Scheme} \label{B_equations}
For the sake of brevity and clarity, the normalizations to the ion parameters for different quantities are adopted.
Time, length, velocity and density are normalized by $\omega_{p_i}^{-1}$, $\lambda_{D_i}$, $v_{th_i}$ and $n_{i0}$ accordingly.
Where $\omega_{pi}  = \sqrt{\frac{n_{i0} e^2}{m_i \epsilon_0}}$, 
$\lambda_{Di} = \sqrt{\frac{\epsilon_0 K_B T_i}{n_{i0} e^2} }$ 
and $v_{th_i} = \sqrt{\frac{K_B T_i}{m_i}}$
are plasma frequency, Debye length and thermal speed of ions respectively ($K_B$ is Boltzmann's constant).
The normalized Vlasov-Poisson set of equations is as follow:
\begin{multline}
\frac{\partial f_s(x,v,t)}{\partial t} 
+ v \frac{\partial f_s(x,v,t)}{\partial x} 
\\ +  \frac{q_s}{m_s} E(x,t) \frac{\partial f_s(x,v,t)}{\partial v} 
= 0, \ \ \  s = i,e
\label{Vlasov}
\end{multline}
\begin{equation}
\frac{\partial^2 \phi(x,t)}{\partial x^2}  = n_e(x,t) - n_i(x,t)
\label{Poisson}
\end{equation}
where $s = i,e$ represents the corresponding species.
The variable $v$ denotes velocity in phase space.
$q_s$ and $m_s$ are normalized by $e$ and $m_i$ respectively.
Densities of the plasma components are calculated through an integration as:
\begin{equation}
n_s(x,t) = n_{0s}\int f_s(x,v,t) dv
\label{density}
\end{equation}
The equilibrium values $n_{s0}$ are assumed to satisfy 
the quasi-neutrality condition ($n_{e0} = n_{i0} $) at the initial step.

In order to introduce a large compressional pulse into the simulation box, 
the Schamel distribution function is used as the initial distribution function. 
\begingroup\makeatletter\def\f@size{8.3}\check@mathfonts
\def\maketag@@@#1{\hbox{\m@th\large\normalfont#1}}%
\begin{equation*}
f_{s}(v) =  
  \left\{\begin{array}{lr}
     A \ \exp \Big[- \big(\sqrt{\frac{\xi_s}{2}} v_0 + \sqrt{\varepsilon(v)} \big)^2 \Big]   &\textrm{if}
      \left\{\begin{array}{lr}
      v<v_0 - \sqrt{\frac{2\varepsilon_{\phi}}{m_s}}\\
      v>v_0+\sqrt{\frac{2\varepsilon_{\phi}}{m_s}} 
      \end{array}\right. \\
     A \ \exp \Big[- \big(\frac{\xi_s}{2} v_0^2 + \beta_s \varepsilon(v) \big) \Big] &\textrm{if}  
     \left\{\begin{array}{lr}
      v>v_0-\sqrt{\frac{2\varepsilon_{\phi}}{m_s}} \\
      v<v_0 + \sqrt{\frac{2\varepsilon_{\phi}}{m_s}} 
      \end{array}\right.
\end{array}\right.
\label{Schamel_Dif}
\end{equation*}\endgroup
in which $A = \sqrt{ \frac{\xi_s}{2 \pi}} n_{0s}$,
and $\xi_s = \frac{m_s}{T_s}$ are amplitude and normalization factors respectively.
$\varepsilon(v) = \frac{\xi_s}{2}(v-v_0)^2 + \phi\frac{q_s}{T_s }$ represents 
the (normalized by $K_B T_i$) energy of particles.
$v_0$ stands for the velocity of the IA soliton.
It is proven that this distribution function 
satisfies the continuity and positiveness conditions
while producing a hole in its phase space \cite{schamel_1,schamel_2}.
We have initialized the Schamel distribution function 
with a stationary perturbation ($v_0 =0$) at $x_0 = 512$ with the form 
\begin{equation}
 \phi = \psi \ exp (\frac{x-x_0}{\Delta})^2,
\end{equation}
in which $\psi$ and  $\Delta$ are the amplitude and width of the initial compressional plus respectively.

In the simulations presented here, 
all the three improvements to the KdV theory
to overcome the disparities are included.
The simulation method employs the Vlasov equation directly, hence all the perturbation orders are self-consistently incorporated.
The limitation on the electron dynamics, as the Boltzmann's fluid, is removed, since electrons dynamics are followed in phase space.
The approximation of cold ions is also removed,
because the dynamics of the ions is treated based on the Vlasov equation.

The Vlasov simulation approach, adopted here, 
has been developed by the authors based on following the trajectories in phase space \cite{nunn1993novel,jenab2011preventing}. 
The simulation method preserves conservation laws for different quantities, 
such as entropy and energy and the distribution function positiveness. 
For all the simulations presented here, the deviation 
from the conservation laws are constantly checked to stay below one percent. 

The constant parameters which remain fixed through all
of our simulations include: 
mass ratio $\frac{m_i}{m_e} = 100$,
time step $d\tau = 0.01$,
temperature ratio $\frac{T_e}{T_i} = 64$
and $L = 4096$, 
where L is the length of the simulation box.
The velocity domain $(v_{min},v_{max}) = (-100,100)$ for electrons and $(-6,6)$ for ions.
They are either $\psi = 0.05$ and  $\Delta = 10$ (small pulse) or $\psi = 0.2$ and  $\Delta = 500$ (large pulse).
The value of trapping parameter is $\beta = -0.1$ for a hole and $\beta = 0.2$ for a hump.
We have considered a two-dimensional phase space with one spatial and one velocity axes.
The phase space grid $(N_x, N_v)$ size is $(4096, 4000)$. 
Hence the grid spacing on spatial and velocity axes are $\delta x = 1.0$,
$\delta v = 0.05$ for electrons and $\delta = 0.003$ for ions.
The periodic boundary condition is employed on the x-direction.

\section{Results and Discussion} \label{Results}

\subsection{Summary of electron hole merging theory} \label{types_collisions}
Ghizzo \textit{et al.} \cite{ghizzo1987bgk} studied the behavior of electron holes generated by the BGK modes
employing direct integration of the Vlasov-Poisson system describing the stability of BGK equilibrium in phase space.
They have shown two main phenomena about electron holes.
Firstly, they tend to behave as quasi-particles in phase space, 
carrying their own inertia and  can be influenced by force fields\cite{ghizzo2014nonlinear}.
This inertia originates from the number of trapped particles inside the electron hole,
hence relates to its size\cite{dupree1983growth}.
As the trapping parameter $\beta$ increases, the inertia of the electron hole grows along with it.
Secondly, before the coalescence they have a central force behavior
which originates from the collective charge they possess.
Therefore, electron holes surface as charged flexible clouds (shielded by the background plasma) in the phase space with their own inertia.
Central force is repulsive in case of IA solitons, since they carry the same kind of charge.

Krasovsky \textit{et al.} \cite{krasovsky1999interaction,krasovsky1999interaction_small,krasovsky2003electrostatic}
have suggested a dissipative process during collision of two holes, based on the energy conservation principle.
The internal energy of a hole 
(kinetic energy of the trapped electrons in the co-moving frame)
grows during collisions, and hence holes warm up. 
This is an irreversible process and causes an effective friction in the energy balance.
In the other words, the dynamics of electron holes collisions resembles a damping oscillator 
with $\epsilon = l/\lambda$ 
playing the role of a damping parameter
($\lambda$ and $l$ are typical distances of charge shielding 
by background plasma and the size of the electron hole respectively).
As the size ($l$) grows the collision becomes more inelastic (for details see Ref. \cite{krasovsky1999interaction_small}).
For two holes to merge, the relative velocity of the holes 
should be slow enough so that the trapped electrons oscillate at least once during collision 
(so called \textit{``merging condition''}). 
In other words, this means they should have enough overlap in velocity.

Three different types of collisions are possible 
based on the damping parameter ($\epsilon$) \cite{krasovsky2003electrostatic}:
\begin{enumerate}
 \item First type ($\epsilon \ll 1 , \ \epsilon \neq 0 $) (bending): \\
Just a simple bending of trajectories in the phase space happens due to repulsion.
This collision happens for holes which do not fulfill the merging condition. 
In other words, they do not overlap in the velocity direction.
\item Second type ($\epsilon < 1, \ \epsilon \neq 0 $) (bending, rotations, merging):\\
Two holes starts bending and then rotate around each other.
During their rotations, they slowly lose their energy and finally merge.
\item Third type ($\epsilon \cong 1$) (bending, merging): \\ 
Two holes rotate once and merge.
\end{enumerate}

\subsection{Benchmarking of the simulation code}
\begin{figure}
 \subfloat{\includegraphics[width=0.5\textwidth]{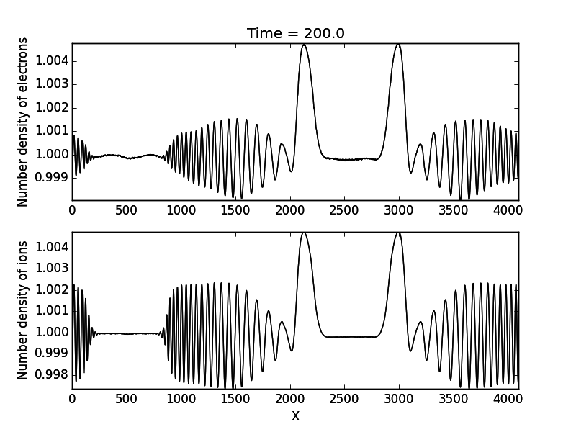} } 
 \caption{Two small self-consistent IA solitons with $\beta=-0.1$, which are created by chain formation process,
 are shown before their head-on collision.
 Each of the IA solitons are followed by an IA wavepacket,
 which are produced by initial breaking of the stationary pulse.
 IA solitons appear as a compressional pulse, 
 while IA wavepacket can be recognized by its amplitude swinging between positive and negative values.}
 \label{fig_small_200}
\end{figure}
As the initial step, a stationary large compressional pulse ($v_0=0$) has been introduced into the simulation domain. 
This pulse breaks into two moving pulses dues to the symmetry in the velocity direction \cite{saeki1998electron}. 
Each of the small/large moving pulse disintegrates into one/number of IA soliton(s). 
During the disintegration, Langmuir or IA wavepacket emerge (see Fig.\ref{fig_small_200}).
This process and the dependency of IA solitons on the initial conditions is discussed in 
Ref.\cite{jenab2016IASWs} extensively. 
\begin{figure*}
 \centering
 \subfloat{\includegraphics[width=0.85\textwidth]{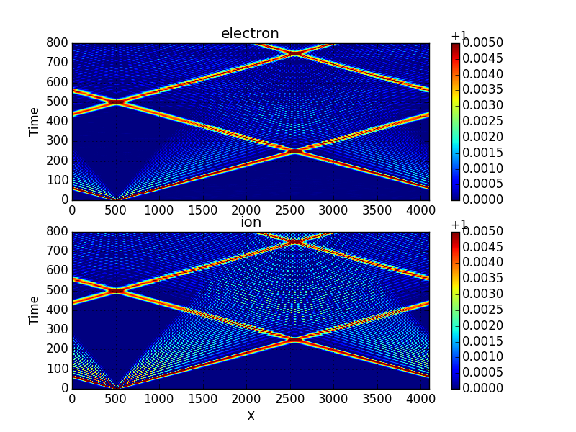} } 
 \caption{Long term simulation results of two small self-consistent IA solitons (with $\beta= 0$)
 in a periodic system are presented.
 Successive mutual collisions between the two solitons happens three times without changing their features.
 Propagation of two IA wavepackets and their collision with the IA solitons are also visible.}
 \label{fig_small_small_long}
\end{figure*}
The stability of the IA solitons against successive mutual collisions 
is presented in Fig.\ref{fig_small_small_long}. 
Based on fully kinetic approach, the stability 
has been discussed throughly in Refs. \cite{jenab2017overtaking, jenab2017pre}.


\subsection{The first type of collisions: bending}
\begin{figure*}
 \subfloat{\includegraphics[width=0.35\textwidth]{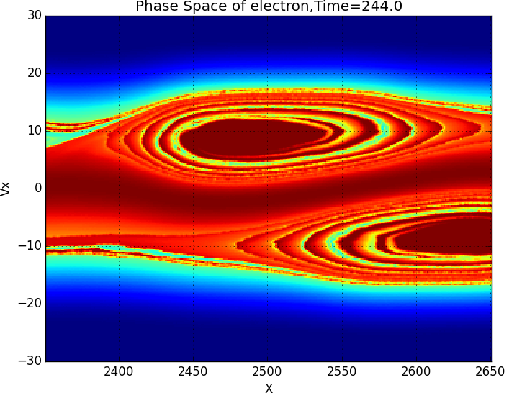} } 
 \subfloat{\includegraphics[width=0.35\textwidth]{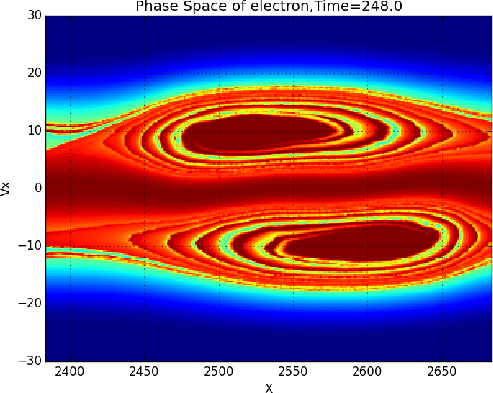} }
 \subfloat{\includegraphics[width=0.35\textwidth]{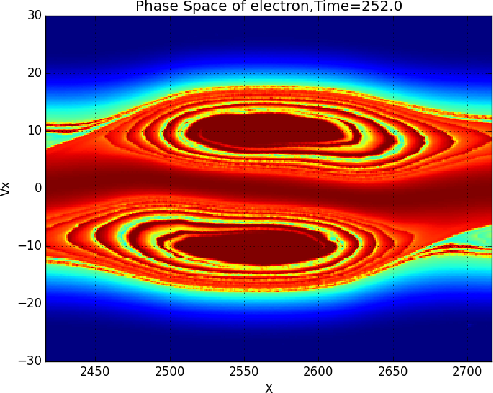} }
 \caption{The collision of two small holes ($\beta= 0$) is presented in the electron phase space, 
 for $\tau= 244$, $248$ and $252$ (starting from the top left corner).
 Due to their high relative velocity (compared to their size in the velocity direction),
 they don't merge (first type of collision).
 However, the \textit{bending} of both their trajectories and shapes can be observed,
 due to the repulsive central force between them. }
 \label{fig_small_small}
\end{figure*}

\begin{figure*}
 \subfloat{\includegraphics[width=0.35\textwidth]{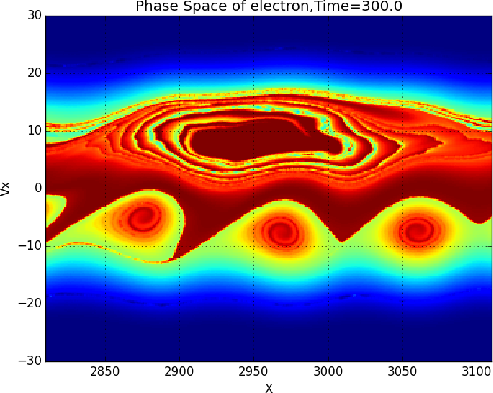} }
 \subfloat{\includegraphics[width=0.35\textwidth]{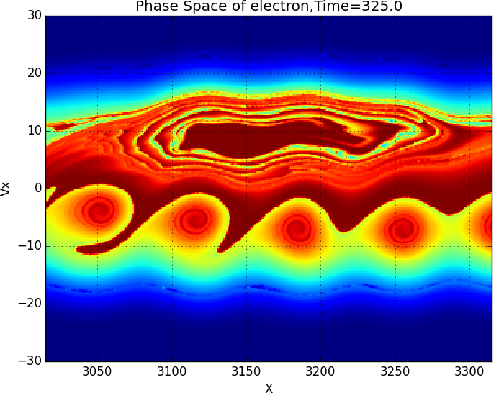} } 
 \subfloat{\includegraphics[width=0.35\textwidth]{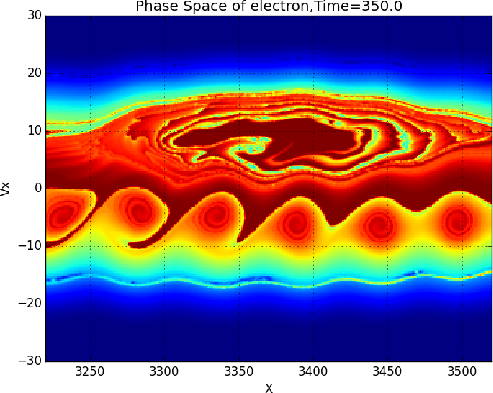} } \\
 \subfloat{\includegraphics[width=0.35\textwidth]{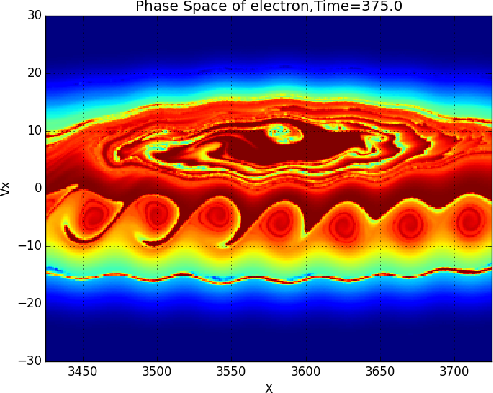} } 
 \subfloat{\includegraphics[width=0.35\textwidth]{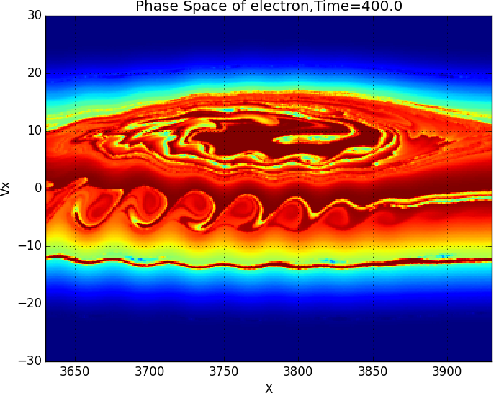} }
 \subfloat{\includegraphics[width=0.35\textwidth]{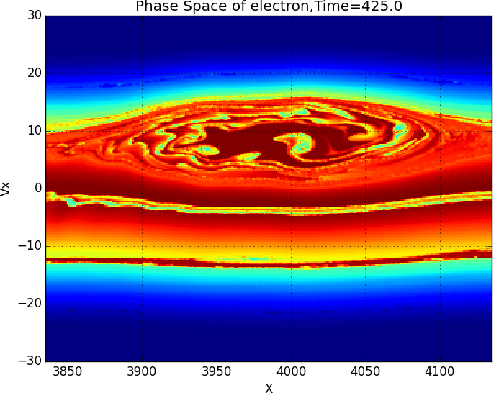} }
 \caption{Successive collisions between one small hole 
 (accompanying an IA soliton, propagating to the right with $\beta = 0$) 
 with number of holes (accompanying IA wavepacket, propagating to the left) 
 are shown in the phase space of electrons (first type of collision).
 Frames belong to time $\tau = 300$, $325$, $350$, $375$, $400$, $425$
 starting from the top left corner. 
 Numerous successive bending disturb 
 the internal structure of the hole (accompanying an IA soliton).}
 \label{fig_small_BGK}
\end{figure*}

The collision of two small IA solitons in the phase space are shown in Fig.\ref{fig_small_small},
which are moving oppositely with the same speed.
Their relative velocity is much higher than their size in velocity-direction of the phase space,
and hence they don't fulfill the merging condition
Therefore, their collision is the first type.
When they come close to each other, background shielding weakens and hence they \textit{feel} each others' presence.
So, their trajectories bend in the opposite direction, representing acceleration due to the repulsion.
Meanwhile,
due to their flexibility,
this bending affects their shapes partially.
They deviate from their bell-shaped structure,
which is predicted by the Schamel distribution function.
After the collision, they continue the temporal propagation without any further interactions with each other.
However, the bending of the shape affects the holes in the later time, since it has introduced some disturbance in the internal structure of the hole. 
As the simulation continues, numerous successive collisions of the IA electron holes with the holes created by IA wavepackets take place. 
Fig.\ref{fig_small_BGK} presents the effect of these collisions on the shape and structure of the IA electron hole. 
Successive bending disturbs the internal structure of the IA electron hole while its shape stays the same.

\subsection{The third type of collisions: bending, merging and splitting}
\begin{figure*}
 \subfloat{\includegraphics[width=0.35\textwidth]{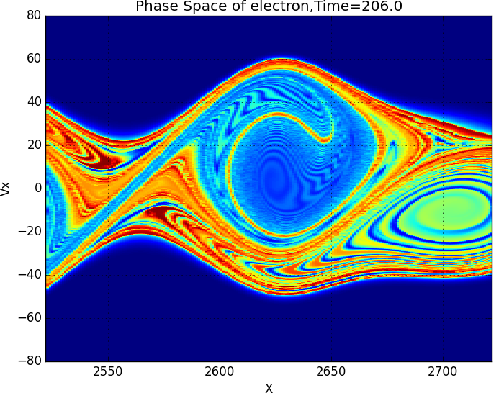} }
 \subfloat{\includegraphics[width=0.35\textwidth]{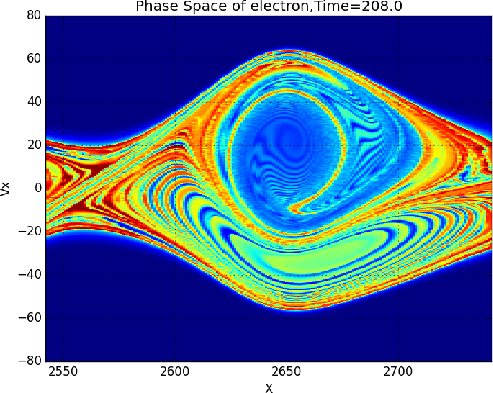} } 
 \subfloat{\includegraphics[width=0.35\textwidth]{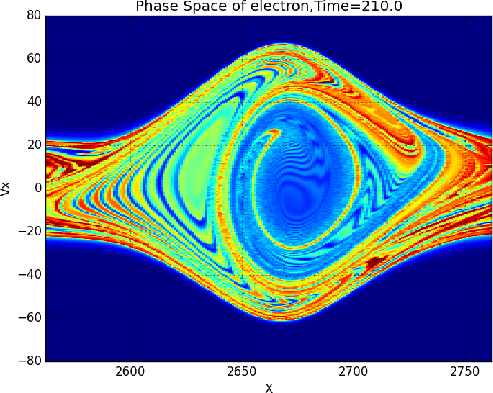} } \\
 \subfloat{\includegraphics[width=0.35\textwidth]{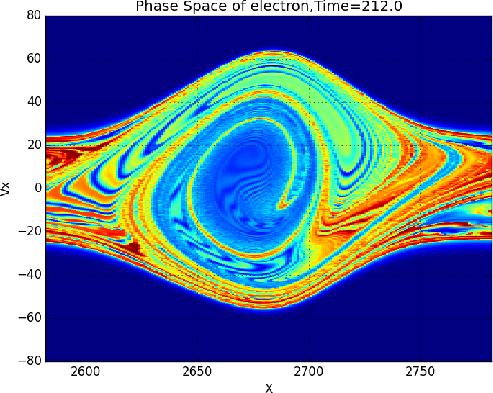} } 
 \subfloat{\includegraphics[width=0.35\textwidth]{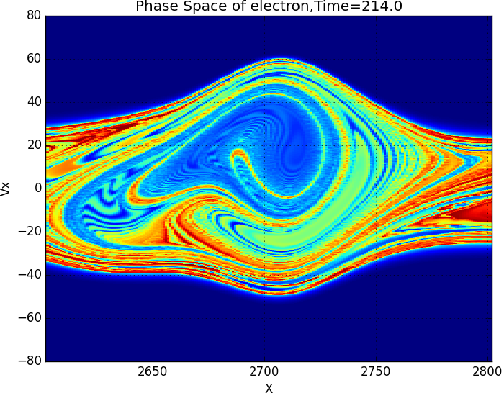} }
 \subfloat{\includegraphics[width=0.35\textwidth]{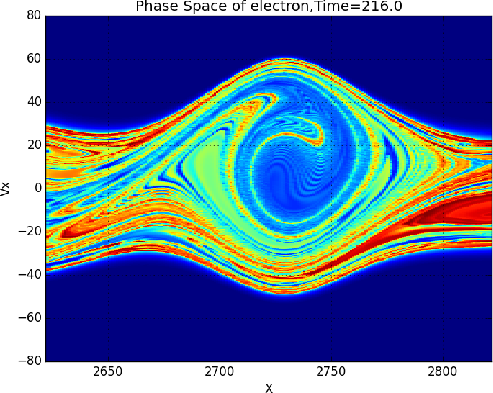} }
 \caption{Collision between a small ($\beta = 0.2$, propagating to left) and 
 a large hole ($\beta = -0.1$, propagating to right) accompanying IA solitons
 is presented in the phase space of electrons. 
 The temporal evolution is shown
 for different time intervals $\tau = 206$, $208$, $210$, $212$, $214$ and $216$ (starting from the top left corner).
 Bending can be witnessed for $\tau = 206$, $208$. 
 The merging starts from $\tau = 210$ as the small hole starts rotating around the big hole. 
 After just one rotation, the accumulated hole split, 
 and the new small hole leave it from the left side, as shown in $\tau = 214$.}
 \label{fig_small_large}
\end{figure*}
Fig. \ref{fig_small_large} presents a collision of a small versus a large hole. 
The temporal progression of the collision follows the theoretical expectation of the third type of collisions. 
The central repulsive force pushes the two holes to bend and then rotate around their collective center of mass. 
However, as it is the case in the classical dynamics,
the large hole appears to stay still, because of its higher share of mass. 
While the small hole seems to rotate around it.
In other words, the center of collective mass is close to the center of mass of the large hole.
On the other hand, due to the flexibility of the holes, the lighter one bends more
(stretch out around the heavier hole), 
and hence the merging seems as if the large hole traps the small one.

After one rotation of holes around each other, on the fluid-level, 
two solitons start departing each other,
hence each of them takes their own share of the trapped particles. 
Therefore, the accumulated population of trapped electrons (i.e. single merged hole) splits into two holes.
Conclusively, the collision has one more step (\textit{splitting}) after the merging to overcome the discrepancy.
Each of the new holes posses the same shape as the old ones (before collisions) accompanying the same IA soliton.
However, their internal structures differ from the old ones.
The splitting of the single merged hole is in contrast to the temporal progression of collisions in case of BGK modes or two-beam instabilities. 
In these cases, single holes have been reported to be stable (if it is the only one in the simulation domain) or tend to merge with other holes,
but they never split.

\begin{figure*}
 \subfloat{\includegraphics[width=0.35\textwidth]{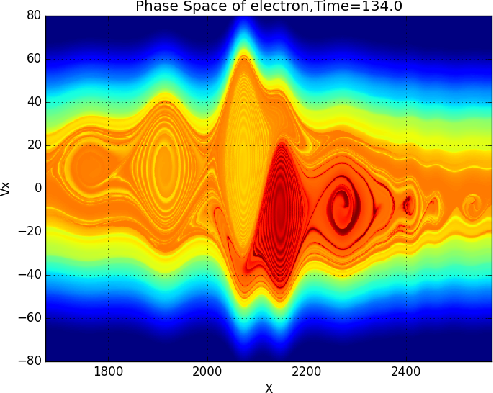} } 
 \subfloat{\includegraphics[width=0.35\textwidth]{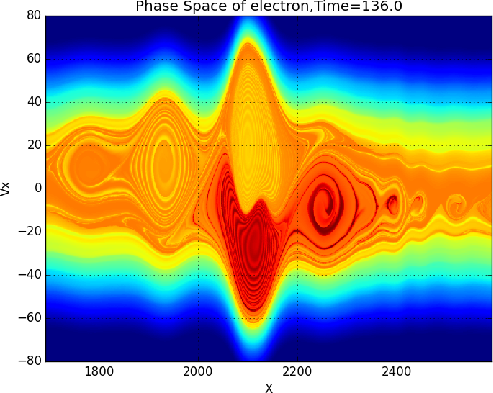} } 
 \subfloat{\includegraphics[width=0.35\textwidth]{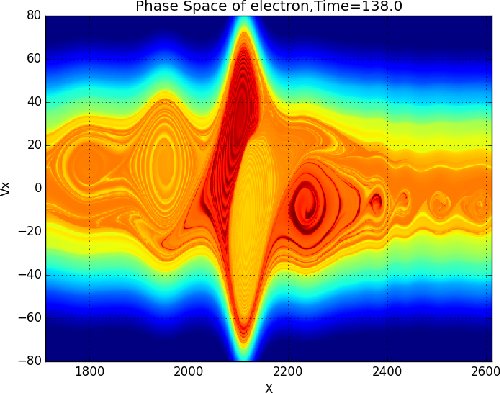} } \\
 \subfloat{\includegraphics[width=0.35\textwidth]{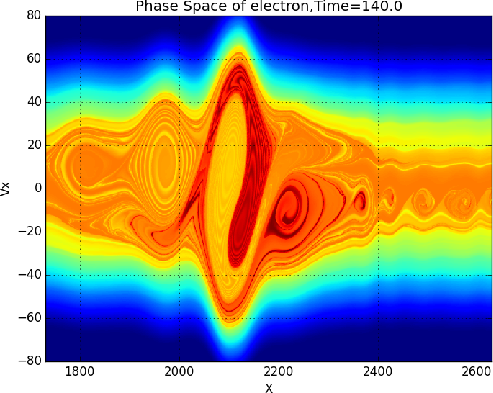} } 
 \subfloat{\includegraphics[width=0.35\textwidth]{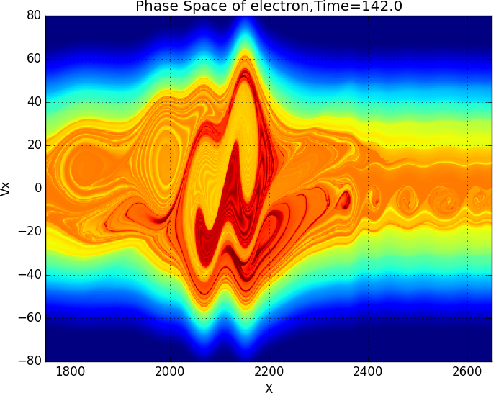} } 
 \subfloat{\includegraphics[width=0.35\textwidth]{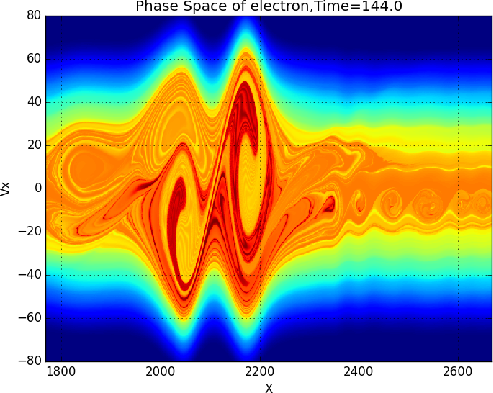} }
 \caption{Collision between a large hollow ($\beta = -0.1$, yellow/bright color, propagating to the right) and 
 a large hump ($\beta = 0.2$ red/dark color, propagating to the left) 
 accompanying IA solitons is shown in the phase space of electrons.
 Temporal evolution is presented for times including 
 $\tau =134$, $136$, $138$, $140$ and $142$ starting from the top left corner.
 The first row of figures represents the bending, taking place in the early stage of collision.
 Furthermore, the holes continue to bend and form a rotation presented in the second row.
 Third row displays the splitting process,
 and also the change in the internal structure of the new hole compared to the old holes.}
 \label{fig_large_large}
\end{figure*}

The dynamics in case of the collision of the two large/heavy holes follows the same pattern as the previous simulation,
and includes bending, merging and splitting (see Fig. \ref{fig_large_large}).
Here, in order to recognize the difference between trapped electron populations (propagating oppositely),
we have shown the collision between a hump (positive trapping parameter $\beta = 0.2$ ) 
and a hole (negative trapping parameter $\beta=-0.1$). 
By checking the electric charge profile of the system, it is confirmed that both of them 
(the hole and the hump) carry the same charge.
Therefore, the central force between them is proved to be repulsive.
Due to the similar size of the two nonlinear structures in the collision,
there exists a symmetry between the two new nonlinear structures.

The  collisions of heavier IA solitons up to $\beta=10$ are also studied and the same processes have been observed.
This suggests that IA solitons,
despite the inertia and strength of central force of holes accompanying them,
survive the collision.
In other words, fluid theory predictions of KdV solitons should be refined 
to overcome the experimental discrepancies concerning the propagation of IA solitons. 
However, the prediction about the stability of KdV solitons against mutual collisions stays true 
even under fully kinetic approach which incorporate all kinetic level refinements,
such as three major developments described in the introduction section.

\subsection{The aftermath of collisions (on kinetic level): Spiral path}
\begin{figure*}
 \begin{tabular}{cc}
  \subfloat{\includegraphics[width=0.3\textwidth]{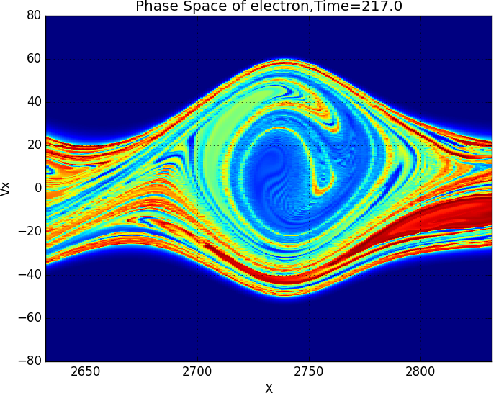} }  
 \subfloat{\includegraphics[width=0.3\textwidth]{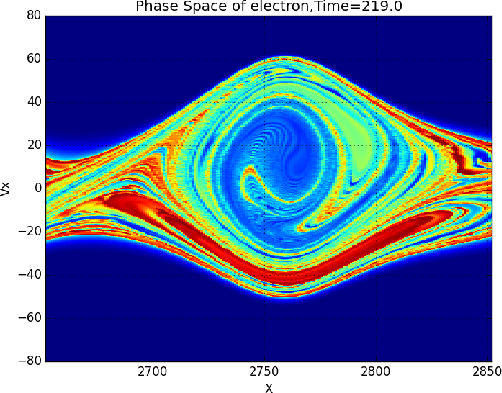} } &
 \multirow{2}{*}[1.0in]{\includegraphics[width=0.42\textwidth]{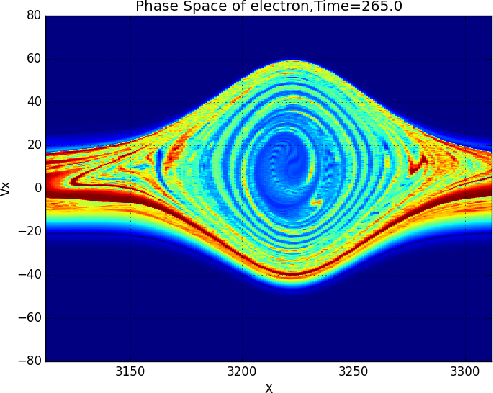}}\\
 \subfloat{\includegraphics[width=0.3\textwidth]{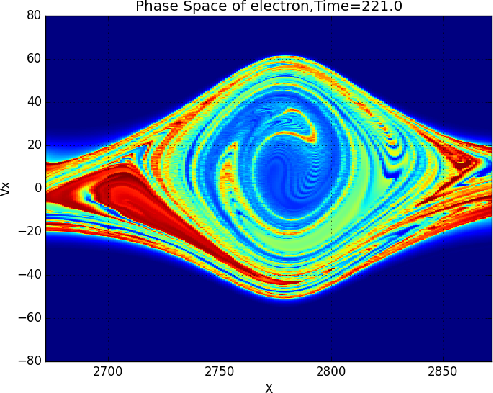} } 
 \subfloat{\includegraphics[width=0.3\textwidth]{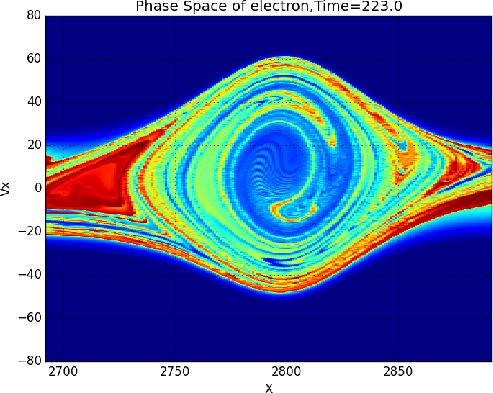} } & {}  
 \end{tabular}
 \caption{Kinetic dynamics of the large hole is presented just 
 after a collision between a large and a small hole. 
 Blue/green color represents the population of the large/small old hole. 
 The population of the small old hole are trapped inside the new large hole.
 One period of rotation is shown from $\tau = 217$ up to $\tau = 223$ (starting from the top left corner).
 After enough rotation, the spiral path can be observed at later time $\tau = 265$.}
 \label{fig_trapping}
\end{figure*}

Following the kinetic dynamics of the holes after the splitting shows that inside each of the holes,
the newly added trapped population of electrons follows a spiral path\cite{bertrand1990nonperiodic} (see Fig.\ref{fig_trapping}). 
\begin{figure*}
 \begin{tabular}{cc}
  \subfloat{\includegraphics[width=0.3\textwidth]{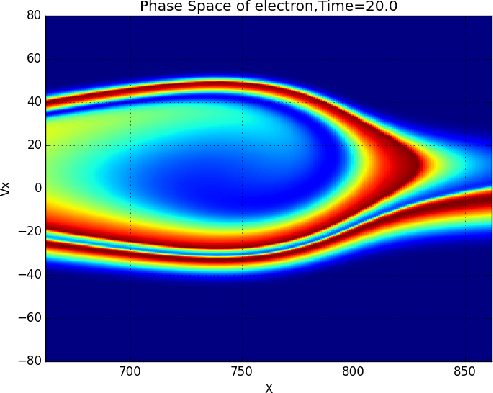} }  
 \subfloat{\includegraphics[width=0.3\textwidth]{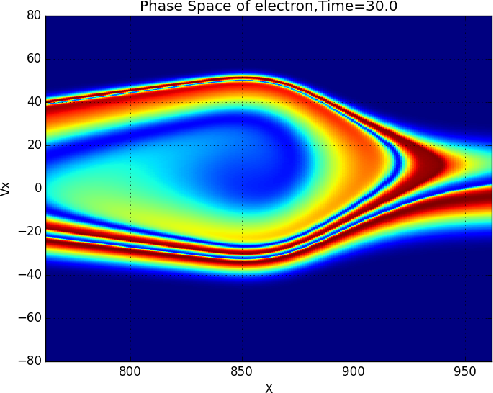} } &
 \multirow{2}{*}[1.0in]{\includegraphics[width=0.42\textwidth]{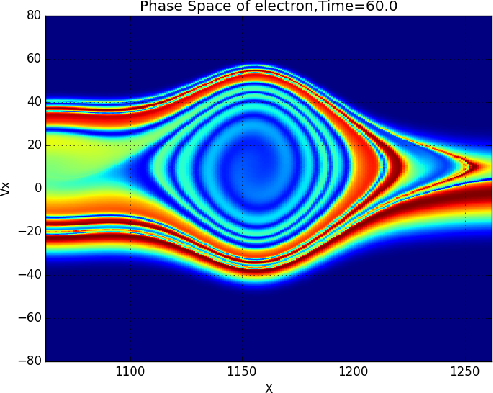}}\\
 \subfloat{\includegraphics[width=0.3\textwidth]{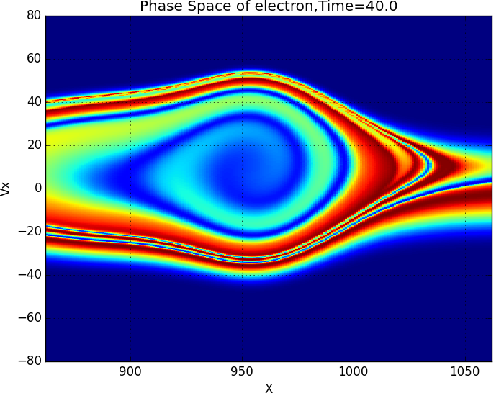} } 
 \subfloat{\includegraphics[width=0.3\textwidth]{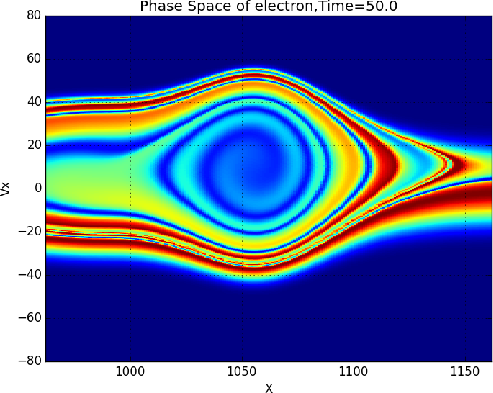} } & {}  
 \end{tabular}
 \caption{Initial steps of an electron hole formation is presented. 
 The trapped population of green and blue colors rotate around each other and go through a spiral path,
 shown at different time intervals, e.g. $\tau =20$, $30$, $40$, $50$, $60$.}
 \label{fig_form}
\end{figure*}
This spiral path can be witnessed in the early stage of the formation of the electron holes
before the collision.
The early stage evolution of an electron hole is shown in Fig.\ref{fig_form}.
This displays how the IA soliton traps the electron population around its vicinity in the phase space.
Therefore, the newly added electron population is trapped 
by the hole and follows the same path as the initial trapped population.

The spiral path of the trapped electrons inside the hole can be seen 
as the continuation of the rotation caused by the central force. 
\textbf{The diameter of the circles decreases }
because of the same damping effect that trapped the population at the beginning,
i.e. irreversible dissipative process in the energy balance. 
Each time the newly added population rotates the center of mass of the electron hole,
it can be considered as a collision between the newly added population and the electron hole.

\section{Conclusions} \label{Conclusions}
The process of head-on collisions of 
two trapped electron populations
accompanying ion-acoustic (IA) solitons
is studied on kinetic level in details for different sizes, 
based on a fully kinetic simulation approach.
In summary, four main findings are reported here. 

(1) The kinetic theory predicts three types of collisions for electron holes. 
Here, the first (bending) and the third (bending, merging) types are observed 
but not the second type (bending, rotations, merging). 
The second type of collision requires IA solitons to bounce around their center of collective mass
which in the phase space manifests as two holes rotating around each other more than once.
This behavior of IA solitons is against the N-solitons solution proposed by fluid theory, 
suggesting that each IA soliton is the normal mode of the system, hence they should not try to accumulate. 
Hence, there is a discrepancy between merging tendency of holes and stability of IA solitons.
Our kinetic simulation shows 
that the later overrule the first tendency 
and the second type collision does not happen.
This conclusion is limited by the temperature ratio and the features of the initial IDP, which is used in these simulations.

(2) In case of the first type collision (bending),
long-lasting consequences of collision 
on the internal structure of the electron hole are reported here,
especially in case of successive collisions.

(3) It is shown that the \textit{splitting} of the single merged hole is the way a plasma chooses to overcome the discrepancy
between stability of IA solitons and the merging tendency of electron holes in the third-type collision.
Hence, a single hole is not stable,
which contrasts with what has been witnessed in case of BGK mode and beam-instabilities.
Simulations with much heavier electron holes ($\beta= 6$ or $10$) 
have shown stability against collisions,
and displayed the same process for the third-type collision.

(4) After splitting, 
each of the new holes has a combination of the trapped electron populations of the old holes. 
The two distinct populations follow a spiral path inside the electron hole as the temporal evolution continues.  
This spiral path is the continuation of the rotation and the merging process of the two colliding holes after splitting.
The newly added population to the holes continues to act as an independent trapped population of electrons inside the hole.
It loses energy due to the dissipative effect in the energy balance 
because of successive collisions (rotating around the center of mass of the electron hole). 
On the other hand, the same evolution pattern is reported here for the early steps of electron hole formation. 
Therefore, after the collision the IA solitons
treat their newly added trapped population the same was as the one trapped earlier in the formation process.

Conclusively, the nonlinear fluid theory of the IA solitons and the kinetic theory of electron holes are two controlling
theories/driving forces behind the dynamics studied here.
Their confrontation in the collision process is such that the system shows most of electron holes kinetic theory
predictions, e.g. bending, one rotation and merging. 
However, stability of IA solitons impose the splitting and removes/cancels the second type of collision,
i.e. multiple rotations.

\section*{Acknowledgment}
This work is based upon research supported by the National Research Foundation 
and Department of Science and Technology.
Any opinion, findings and conclusions or recommendations expressed in this 
material are those of the authors and therefore the NRF and DST do not accept 
any liability in regard thereto.


\begin{thebibliography}{10}
\providecommand{\url}[1]{#1}
\csname url@samestyle\endcsname
\providecommand{\newblock}{\relax}
\providecommand{\bibinfo}[2]{#2}
\providecommand{\BIBentrySTDinterwordspacing}{\spaceskip=0pt\relax}
\providecommand{\BIBentryALTinterwordstretchfactor}{4}
\providecommand{\BIBentryALTinterwordspacing}{\spaceskip=\fontdimen2\font plus
\BIBentryALTinterwordstretchfactor\fontdimen3\font minus
  \fontdimen4\font\relax}
\providecommand{\BIBforeignlanguage}[2]{{%
\expandafter\ifx\csname l@#1\endcsname\relax
\typeout{** WARNING: IEEEtran.bst: No hyphenation pattern has been}%
\typeout{** loaded for the language `#1'. Using the pattern for}%
\typeout{** the default language instead.}%
\else
\language=\csname l@#1\endcsname
\fi
#2}}
\providecommand{\BIBdecl}{\relax}
\BIBdecl

\bibitem{ikezi1970formation}
H.~Ikezi, R.~Taylor, and D.~Baker, ``Formation and interaction of ion-acoustic
  solitions,'' \emph{Physical Review Letters}, vol.~25, no.~1, p.~11, 1970.

\bibitem{ikezi1973experiments}
H.~Ikezi, ``Experiments on ion-acoustic solitary waves,'' \emph{Research
  report}, vol. 149, pp. 1--36, 1973.

\bibitem{nakamura1996experiments}
Y.~Nakamura and H.~Sugai, ``Experiments on ion-acoustic solitons in a plasma,''
  \emph{Chaos, Solitons \& Fractals}, vol.~7, no.~7, pp. 1023--1031, 1996.

\bibitem{nakamura1982experiments}
Y.~Nakamura, ``Experiments on ion-acoustic solitons in plasmas invited review
  article,'' \emph{IEEE Transactions on Plasma Science}, vol.~10, no.~3, pp.
  180--195, 1982.

\bibitem{Washimi1966996}
H.~Washimi and T.~Taniuti, ``Propagation of ion-acoustic solitary waves of
  small amplitude,'' \emph{Physical Review Letters}, vol.~17, no.~19, p. 996,
  1966.

\bibitem{kakutani1968reductive}
T.~Kakutani, H.~Ono, T.~Taniuti, and C.-C. Wei, ``Reductive perturbation method
  in nonlinear wave propagation ii. application to hydromagnetic waves in cold
  plasma,'' \emph{Journal of the Physical Society of Japan}, vol.~24, no.~5,
  pp. 1159--1166, 1968.

\bibitem{Gardner19671095}
C.~S. Gardner, J.~M. Greene, M.~D. Kruskal, and R.~M. Miura, ``Method for
  solving the korteweg-devries equation,'' \emph{Physical Review Letters},
  vol.~19, no.~19, p. 1095, 1967.

\bibitem{taniuti1968reductive}
T.~Taniuti and C.-C. Wei, ``Reductive perturbation method in nonlinear wave
  propagation. i,'' \emph{Journal of the Physical Society of Japan}, vol.~24,
  no.~4, pp. 941--946, 1968.

\bibitem{Zabusky1965}
N.~J. Zabusky and M.~D. Kruskal, ``Interaction of" solitons" in a collisionless
  plasma and the recurrence of initial states,'' \emph{Physical review
  letters}, vol.~15, no.~6, p. 240, 1965.

\bibitem{schamel_4}
H.~Schamel, ``A modified korteweg-de vries equation for ion acoustic wavess due
  to resonant electrons,'' \emph{Journal of Plasma Physics}, vol.~9, no.~03,
  pp. 377--387, 1973.

\bibitem{Kakad2013}
A.~Kakad, Y.~Omura, and B.~Kakad, ``Experimental evidence of ion acoustic
  soliton chain formation and validation of nonlinear fluid theory,''
  \emph{Physics of Plasmas (1994-present)}, vol.~20, no.~6, p. 062103, 2013.

\bibitem{Sharma2015}
S.~Sharma, S.~Sengupta, and A.~Sen, ``Particle-in-cell simulation of large
  amplitude ion-acoustic solitons,'' \emph{Physics of Plasmas (1994-present)},
  vol.~22, no.~2, p. 022115, 2015.

\bibitem{Kakad20145589}
B.~Kakad, A.~Kakad, and Y.~Omura, ``Nonlinear evolution of ion acoustic
  solitary waves in space plasmas: Fluid and particle-in-cell simulations,''
  \emph{Journal of Geophysical Research: Space Physics}, vol. 119, no.~7, pp.
  5589--5599, 2014.

\bibitem{Qi20153815}
X.~Qi, Y.-X. Xu, X.-Y. Zhao, L.-Y. Zhang, W.-S. Duan, and L.~Yang,
  ``Application of particle-in-cell simulation to the description of ion
  acoustic solitary waves,'' \emph{IEEE Transactions on Plasma Science},
  vol.~43, no.~11, pp. 3815--3820, 2015.

\bibitem{jenab2017overtaking}
S.~M. Hosseini~Jenab and F.~Spanier, ``Simulation study of overtaking of
  ion-acoustic solitons in the fully kinetic regime,'' \emph{Physics of
  Plasmas}, vol.~24, no.~3, p. 032305, 2017.

\bibitem{jenab2017pre}
S.~M. Hosseini~Jenab and F.~Spanier, ``Fully kinetic simulation study of ion-acoustic solitons in the
  presence of trapped electrons,'' \emph{Physical Review E}, vol.~95, no.~5, p.
  053201, 2017.

\bibitem{franz1998polar}
J.~R. Franz, P.~M. Kintner, and J.~S. Pickett, ``Polar observations of coherent
  electric field structures,'' \emph{Geophysical research letters}, vol.~25,
  no.~8, pp. 1277--1280, 1998.

\bibitem{matsumoto1994electrostatic}
H.~Matsumoto, H.~Kojima, T.~Miyatake, Y.~Omura, M.~Okada, I.~Nagano, and
  M.~Tsutsui, ``Electrostatic solitary waves (esw) in the magnetotail: Ben wave
  forms observed by geotail,'' \emph{Geophysical Research Letters}, vol.~21,
  no.~25, pp. 2915--2918, 1994.

\bibitem{kojima1997geotail}
H.~Kojima, H.~Matsumoto, S.~Chikuba, S.~Horiyama, M.~Ashour-Abdalla, and
  R.~Anderson, ``Geotail waveform observations of broadband/narrowband
  electrostatic noise in the distant tail,'' \emph{Journal of Geophysical
  Research: Space Physics}, vol. 102, no.~A7, pp. 14\,439--14\,455, 1997.

\bibitem{catte1151998fast}
C.~Catte115, D.~Klumparfi, E.~Shelley, W.~Petersonfi, E.~Moebius, and
  L.~Kistler, ``Fast satellite observations of large-amplitude solitary
  structures,'' \emph{Geophysical Research Letters}, vol.~25, no.~12, pp.
  2041--2044, 1998.

\bibitem{pickett2003solitary}
J.~Pickett, J.~Menietti, D.~Gurnett, B.~Tsurutani, P.~Kintner, E.~Klatt, and
  A.~Balogh, ``Solitary potential structures observed in the magnetosheath by
  the cluster spacecraft,'' \emph{Nonlinear Processes in Geophysics}, vol.~10,
  no. 1/2, pp. 3--11, 2003.

\bibitem{hobara2008cluster}
Y.~Hobara, S.~Walker, M.~Balikhin, O.~Pokhotelov, M.~Gedalin,
  V.~Krasnoselskikh, M.~Hayakawa, M.~Andr{\'e}, M.~Dunlop, H.~R{\`e}me
  \emph{et~al.}, ``Cluster observations of electrostatic solitary waves near
  the earth's bow shock,'' \emph{Journal of Geophysical Research: Space
  Physics}, vol. 113, no.~A5, 2008.

\bibitem{pickett2004solitary}
J.~Pickett, S.~Kahler, L.-J. Chen, R.~Huff, O.~Santolik, Y.~Khotyaintsev,
  P.~D{\'e}cr{\'e}au, D.~Winningham, R.~Frahm, M.~Goldstein \emph{et~al.},
  ``Solitary waves observed in the auroral zone: the cluster multi-spacecraft
  perspective,'' \emph{Nonlinear Processes in Geophysics}, vol.~11, no.~2, pp.
  183--196, 2004.

\bibitem{tran1979ion}
M.~Tran, ``Ion acoustic solitons in a plasma: A review of their experimental
  properties and related theories,'' \emph{Physica Scripta}, vol.~20, no. 3-4,
  p. 317, 1979.

\bibitem{Tappert19722446}
F.~Tappert, ``Improved korteweg-devries equation for ion-acoustic waves,''
  \emph{Physics of Fluids}, vol.~15, no.~12, pp. 2446--2447, 1972.

\bibitem{tagare1973effect}
S.~Tagare, ``Effect of ion temperature on propagation of ion-acoustic solitary
  waves of small amplitudes in collisionless plasma,'' \emph{Plasma Physics},
  vol.~15, no.~12, p. 1247, 1973.

\bibitem{sakanaka1972formation}
P.~Sakanaka, ``Formation and interaction of ion-acoustic solitary waves in a
  collisionless warm plasma,'' \emph{Physics of Fluids (1958-1988)}, vol.~15,
  no.~2, pp. 304--310, 1972.

\bibitem{tran1976trapped}
M.~Tran and R.~Means, ``Trapped electron in ion acoustic waves and solitons,''
  \emph{Physics Letters A}, vol.~59, no.~2, pp. 128--130, 1976.

\bibitem{schamel_3}
H.~Schamel, ``Stationary solitary, snoidal and sinusoidal ion acoustic waves,''
  \emph{Plasma Physics}, vol.~14, no.~10, p. 905, 1972.

\bibitem{ichikawa1977solitons}
Y.~Ichikawa and S.~Watanabe, ``Solitons, envelope solitons in collisionless
  plasmas,'' \emph{Le Journal de Physique Colloques}, vol.~38, no.~C6, pp.
  C6--15, 1977.

\bibitem{watanabe1978interpretation}
S.~Watanabe, ``Interpretation of experiments on ion acoustic soliton,''
  \emph{Journal of the Physical Society of Japan}, vol.~44, no.~2, pp.
  611--617, 1978.

\bibitem{h1976contribution}
Y.~H.~Ichikawa, T.~Mitsuhashi, and K.~Konno, ``Contribution of higher order
  terms in the reductive perturbation theory. i. a case of weakly dispersive
  wave,'' \emph{Journal of the Physical Society of Japan}, vol.~41, no.~4, pp.
  1382--1386, 1976.

\bibitem{berk1970phase}
H.~Berk, C.~Nielsen, and K.~Roberts, ``Phase space hydrodynamics of equivalent
  nonlinear systems: Experimental and computational observations,''
  \emph{Physics of Fluids (1958-1988)}, vol.~13, no.~4, pp. 980--995, 1970.

\bibitem{omura1996electron}
Y.~Omura, H.~Matsumoto, T.~Miyake, and H.~Kojima, ``Electron beam instabilities
  as generation mechanism of electrostatic solitary waves in the magnetotail,''
  \emph{Journal of Geophysical Research: Space Physics}, vol. 101, no.~A2, pp.
  2685--2697, 1996.

\bibitem{ghizzo1988stability}
A.~Ghizzo, B.~Izrar, P.~Bertrand, E.~Fijalkow, M.~Feix, and M.~Shoucri,
  ``Stability of bernstein--greene--kruskal plasma equilibria. numerical
  experiments over a long time,'' \emph{Physics of Fluids (1958-1988)},
  vol.~31, no.~1, pp. 72--82, 1988.

\bibitem{eliasson2005solitary}
B.~Eliasson and P.~Shukla, ``Solitary phase-space holes in pair plasmas,''
  \emph{Physical Review E}, vol.~71, no.~4, p. 046402, 2005.

\bibitem{krasovsky1999interaction}
V.~Krasovsky, H.~Matsumoto, and Y.~Omura, ``Interaction dynamics of
  electrostatic solitary waves,'' \emph{Nonlinear Processes in Geophysics},
  vol.~6, no. 3/4, pp. 205--209, 1999.

\bibitem{krasovsky1999interaction_small}
V.~L. Krasovsky, H.~Matsumoto, and Y.~Omura, ``Interaction of small phase
  density holes,'' \emph{Physica Scripta}, vol.~60, no.~5, p. 438, 1999.

\bibitem{krasovsky2003electrostatic}
V.~Krasovsky, H.~Matsumoto, and Y.~Omura, ``Electrostatic solitary waves as
  collective charges in a magnetospheric plasma: Physical structure and
  properties of bernstein--greene--kruskal (bgk) solitons,'' \emph{Journal of
  Geophysical Research: Space Physics}, vol. 108, no.~A3, 2003.

\bibitem{ghizzo1987bgk}
A.~Ghizzo, B.~Izrar, P.~Bertrand, M.~Feix, E.~Fijalkow, and M.~Shoucri, ``Bgk
  structures as quasi-particles,'' \emph{Physics Letters A}, vol. 120, no.~4,
  pp. 191--195, 1987.

\bibitem{Hirota1971}
R.~Hirota, ``Exact solution of the korteweg—de vries equation for multiple
  collisions of solitons,'' \emph{Physical Review Letters}, vol.~27, no.~18, p.
  1192, 1971.

\bibitem{Hirota1972}
R.~Hirota, ``Exact solution of the modified korteweg-de vries equation for
  multiple collisions of solitons,'' \emph{Journal of the Physical Society of
  Japan}, vol.~33, no.~5, pp. 1456--1458, 1972.

\bibitem{jenab2016IASWs}
S.~M. Hosseini~Jenab and F.~Spanier, ``Study of trapping effect on ion-acoustic
  solitary waves based on a fully kinetic simulation approach,'' \emph{Physics
  of Plasmas}, vol.~23, no.~10, p. 102306, 2016.

\bibitem{schamel_1}
H.~Schamel, ``Stationary solutions of the electrostatic vlasov equation,''
  \emph{Plasma Physics}, vol.~13, no.~6, p. 491, 1971.

\bibitem{schamel_2}
H.~Schamel, ``Non-linear electrostatic plasma waves,'' \emph{Journal of Plasma
  Physics}, vol.~7, no.~01, pp. 1--12, 1972.

\bibitem{nunn1993novel}
D.~Nunn, ``A novel technique for the numerical simulation of hot collision-free
  plasma; vlasov hybrid simulation,'' \emph{Journal of Computational Physics},
  vol. 108, no.~1, pp. 180--196, 1993.

\bibitem{jenab2011preventing}
H.~Abbasi, S.~M. Hosseini~Jenab, and H.~H. Pajouh, ``Preventing the recurrence
  effect in the vlasov simulation by randomizing phase-point velocities in
  phase space,'' \emph{Physical Review E}, vol.~84, no.~3, p. 036702, 2011.

\bibitem{ghizzo2014nonlinear}
A.~Ghizzo and D.~Del~Sarto, ``Nonlinear nature of kinetic undamped waves
  induced by electrostatic turbulence in stimulated raman backscattering,''
  \emph{The European Physical Journal D}, vol.~68, no.~10, pp. 1--10, 2014.

\bibitem{dupree1983growth}
T.~H. Dupree, ``Growth of phase-space density holes,'' \emph{The Physics of
  fluids}, vol.~26, no.~9, pp. 2460--2481, 1983.

\bibitem{saeki1998electron}
K.~Saeki and H.~Genma, ``Electron-hole disruption due to ion motion and
  formation of coupled electron hole and ion-acoustic soliton in a plasma,''
  \emph{Physical review letters}, vol.~80, no.~6, p. 1224, 1998.

\bibitem{bertrand1990nonperiodic}
P.~Bertrand, A.~Ghizzo, T.~Johnston, M.~Shoucri, E.~Fijalkow, and M.~Feix, ``A
  nonperiodic euler--vlasov code for the numerical simulation of laser--plasma
  beat wave acceleration and raman scattering,'' \emph{Physics of Fluids B:
  Plasma Physics (1989-1993)}, vol.~2, no.~5, pp. 1028--1037, 1990.

\end{thebibliography}
\end{document}